\documentclass[conference]{IEEEtran}
\usepackage{amsmath}
\usepackage{latexsym}
\usepackage{graphicx}
\usepackage{bbding}
\usepackage{indentfirst}
\IEEEoverridecommandlockouts
\begin{document}

\title{Secure Wireless Information and Power Transfer in Large-Scale MIMO Relaying Systems with Imperfect CSI}
\author{\authorblockN{Xiaoming~Chen$^{\dagger,\ddagger}$,
Jian~Chen$^{\dagger}$, and Tao~Liu$^{\dagger}$
\\$^{\dagger}$ College of Electronic and Information Engineering, Nanjing
University of Aeronautics and Astronautics, China.
\\$^{\ddagger}$ National Mobile Communications Research Laboratory, Southeast
University, China.
\\Email: \{chenxiaoming, chenjian04, tliu\}@nuaa.edu.cn
\thanks{This work was supported by the National Natural
Science Foundation of China (No. 61301102), the Natural Science
Foundation of Jiangsu Province (No. BK20130820), the open research
fund of National Mobile Communications Research Laboratory,
Southeast University (No. 2012D16), the Doctoral Fund of Ministry of
Education of China (No. 20123218120022) and the China Postdoctoral
Science Foundation Funded Project (No. 2014T70517).}}} \maketitle

\begin{abstract}
In this paper, we address the problem of secure wireless information
and power transfer in a large-scale multiple-input multiple-output
(LS-MIMO) amplify-and-forward (AF) relaying system. The advantage of
LS-MIMO relay is exploited to enhance wireless security,
transmission rate and energy efficiency. In particular, the
challenging issues incurred by short interception distance and long
transfer distance are well addressed simultaneously. Under very
practical assumptions, i.e., no eavesdropper's channel state
information (CSI) and imperfect legitimate channel CSI, this paper
investigates the impact of imperfect CSI, and obtains an explicit
expression of the secrecy outage capacity in terms of transmit power
and channel condition. Then, we propose an optimal power splitting
scheme at the relay to maximize the secrecy outage capacity.
Finally, our theoretical claims are validated by simulation results.
\end{abstract}

\section{Introduction}
Energy harvesting facilitates the battery charging to prolong the
lifetime of wireless networks, especially under some extreme
conditions, such as battle-field, underwater and body area networks
\cite{Medical}. Wherein, electromagnetic wave based wireless power
transfer has received considerable research attentions from academic
and industry due to two-fold reasons. First, it is a controllable
power transfer mode. Second, information and power can be
simultaneously transferred in the form of electromagnetic wave
\cite{SWIPT1} \cite{SWIPT2}.

Similar to information transmission, wireless power transfer may
suffer from channel fading, resulting in low energy efficiency. In
this end, energy beamforming is proposed by deploying multiple
antennas at the power source \cite{Energybeamforming1}
\cite{Energybeamforming2}. The impact of channel state information
(CSI) at the power source on the performance of wireless information
and power transfer (WIPT) is quantitatively analyzed in
\cite{Energybeamforming3}. Recently, large scale multiple input
multiple output (LS-MIMO) techniques are also introduced to
significantly improve the efficiency of WIPT by exploiting the high
spatial resolution \cite{LS-MIMO}. Moreover, relaying technique is
also proved as an effective way of improving the performance of WIPT
by shortening the transfer distance, since the distance has a great
impact on both information and power transfer \cite{Relay1}. A
two-way relaying scheme is proposed in \cite{Relay2} to offer a
higher transmission rate with the harvested energy. In fact, through
combing relaying schemes and multi-antenna techniques, especially
the LS-MIMO techniques, the performance of WIPT can be improved
significantly, even in the case of long-distance transfer. However,
to the best of our knowledge, there is no work studying the problem
of the LS-MIMO relaying techniques for WIPT.

Meanwhile, information transfer may encounter interception from the
eavesdropper due to the broadcast nature of wireless channels. In
recent years, as a supplementary of encryption techniques, physical
layer security is used to realize secure communications, by
exploiting wireless channel characteristics, i.e., fading and noise.
The performance of physical layer security is determined by the
performance difference between the legitimate and eavesdropper
channels \cite{PLS}. Thus, the LS-MIMO relaying technique is also an
effective way of improving the secrecy performance
\cite{LS-MIMORelaying}. For secure WIPT, long-distance transfer and
short-distance interception are two challenging issues
\cite{SecureWIPT1} \cite{SecureWIPT2}. To solve them, we introduce
the amplify-and-forward (AF) LS-MIMO relaying technique into secure
WIPT. The contributions of this paper are two-fold:

\begin{enumerate}
\item We derive an explicit expression of the secrecy outage capacity
for such a secure WIPT system in terms of transmit power, CSI
accuracy and transfer distance.

\item We propose a power splitting scheme at the relay, so as to
maximize the secrecy outage capacity.
\end{enumerate}

The rest of this paper is organized as follows. We first give an
overview of the secure WIPT system based on the LS-MIMO AF relaying
scheme in Section II, and then derive the secrecy outage capacity
under imperfect CSI in Section III. In Section IV, we present some
simulation results to validate the effectiveness of the proposed
scheme. Finally, we conclude the whole paper in Section V.

\section{System Model}
\begin{figure}[h] \centering
\includegraphics [width=0.4\textwidth] {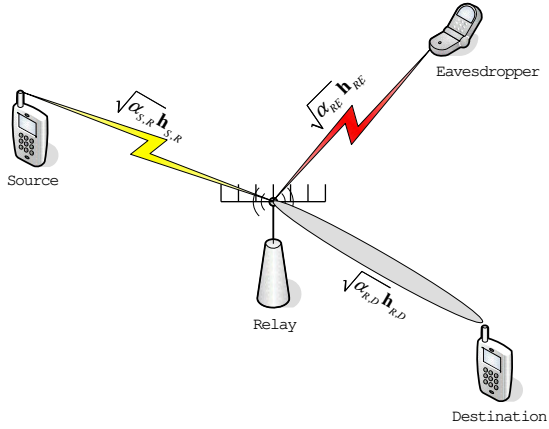}
\caption {An overview of the secure WIPT relaying system.}
\label{Fig1}
\end{figure}

We consider a time division duplex (TDD) LS-MIMO AF relaying system,
where a single antenna source communicates with a single antenna
destination aided by a multi-antenna relay, while a single antenna
passive eavesdropper intends to intercept the message, as shown in
Fig.\ref{Fig1}. Note that the number of antenna at the relay $N_R$
is quite large for such an LS-MIMO relaying system, i.e., $N_R=100$
or greater. The relay only has limited energy to maintain the active
state, so it needs to harvest enough energy from the source for
information transmission. Considering the limited storage, the relay
should be charged from the source slot by slot.

The whole system is operated in slotted time of length $T$, and the
relay works in the half-duplex mode. Then, the information
transmission from the source to the destination via the aid of the
relay requires two time slots. Specifically, in the first time slot,
the source sends the signal, and the relay splits the received
signal into two streams, one for energy harvesting and the other for
information processing. During the second time slot, the relay
forwards the post-processing signal to the destination with the
harvested energy. Note that the direct link from the source to the
destination is unavailable due to a long distance. We assume that
the eavesdropper is far away from the source and is close to the
relay, since it thought the signal comes from the relay. Note that
it is a common assumption in related literature
\cite{LS-MIMORelaying} \cite{SecureRelay}, since it is difficult for
the eavesdropper to overhear the signals from the source and the
relay simultaneously. Then, the eavesdropper only monitors the
transmission from the relay to the destination.

We use $\sqrt{\alpha_{i,j}}\textbf{h}_{i,j}$ to denote the channel
from $i$ to $j$, where $i\in\{S,R\}$ and $j\in\{R,D,E\}$ with $S, R,
D, E$ representing the source, the relay, the destination and the
eavesdropper, respectively. $\alpha_{i,j}$ is the distance-dependent
path loss and $\textbf{h}_{i,j}$ is the small scale fading. In this
paper, we model $\textbf{h}_{i,j}$ as Gaussian distribution with
zero mean and unit variance. $\alpha_{i,j}$ remains constant during
a relatively long period and $\textbf{h}_{i,j}$ fades independently
slot by slot. Thus, the received signal at the relay in the first
time slot can be expressed as
\begin{equation}
\textbf{y}_R=\sqrt{P_S\alpha_{S,R}}\textbf{h}_{S,R}s+\textbf{n}_R,\label{eqn1}
\end{equation}
where $s$ is the normalized Gaussian distributed transmit signal,
$P_S$ is the transmit power at the source, and $\textbf{n}_R$ is the
additive Gaussian white noise with zero mean and unit variance at
the relay. As mentioned above, the relay splits the received signal
$\textbf{y}_R$ into energy harvesting stream with proportional
factor $\theta$ and signal processing stream with $1-\theta$. Then,
according to the law of energy conservation, the harvesting energy
at the relay is given by
\begin{eqnarray}
E_h=\theta\eta\alpha_{S,R}P_S\|\textbf{h}_{S,R}\|^2T,\label{eqn3}
\end{eqnarray}
where the constant parameter $\eta$, scaling from 0 to 1, is the
efficiency ratio at the relay for converting the harvested energy to
the electrical energy to be stored \cite{Energybeamforming2}
\cite{Energybeamforming3}. With the harvested energy, the relay
forwards the post-processing signal $\textbf{r}$ to the destination.
Then the received signals at the destination and the eavesdropper
are given by
\begin{equation}
y_D=\sqrt{\alpha_{R,D}}\textbf{h}_{R,D}^H\textbf{r}+n_D,\label{eqn4}
\end{equation}
and
\begin{equation}
y_{E}=\sqrt{\alpha_{R,E}}\textbf{h}_{R,E}^H\textbf{r}+n_{E},\label{eqn5}
\end{equation}
respectively, where $n_D$ and $n_{E}$ are the additive Gaussian
white noises with zero mean and unit variance at the destination and
the eavesdropper. $\textbf{r}=\sqrt{1-\theta}\textbf{W}\textbf{y}_R$
is the post-processing signal with $\textbf{W}$ being a transform
matrix.

We assume the relay has full CSI $\textbf{h}_{S,R}$ by channel
estimation, and gets partial CSI $\textbf{h}_{R,D}$ via channel
reciprocity in TDD systems. Due to duplex delay between uplink and
downlink, there is a certain degree of mismatch between the
estimated CSI $\hat{\textbf{h}}_{R,D}$ and the real CSI
$\textbf{h}_{R,D}$, whose relation can be expressed as
\cite{CSIMismatch}
\begin{equation}
\textbf{h}_{R,D}=\sqrt{\rho}\hat{\textbf{h}}_{R,D}+\sqrt{1-\rho}\textbf{e},\label{eqn6}
\end{equation}
where $\textbf{e}$ is the error noise vector with independent and
identically distributed (i.i.d.) zero mean and unit variance complex
Gaussian entries. $\rho$, scaling from 0 to 1, is the correlation
coefficient between $\hat{\textbf{h}}_{R,D}$ and $\textbf{h}_{R,D}$.
A larger $\rho$ means better CSI accuracy. If $\rho=1$, the relay
has full CSI $\textbf{h}_{R,D}$. Additionally, due to the hidden
property of the eavesdropper, the CSI $\textbf{h}_{R,E}$ is
unavailable. Therefore, $\textbf{W}$ is designed only based on
$\textbf{h}_{S,R}$ and $\hat{\textbf{h}}_{R,D}$, but is independent
of $\textbf{h}_{R,E}$. Considering the better performance of maximum
ratio combination (MRC) and maximum ratio transmission (MRT) in
LS-MIMO systems, we design $\textbf{W}$ by combining MRC and MRT.
Mathematically, the transform matrix is given by
\begin{equation}
\textbf{W}=\kappa\frac{\hat{\textbf{h}}_{R,D}}{\|\hat{\textbf{h}}_{R,D}\|}\frac{\textbf{h}_{S,R}^H}{\|\textbf{h}_{S,R}\|},\label{eqn7}
\end{equation}
where $\kappa$ is the power constraint factor. To fulfill the energy
constraint at the relay, $\kappa$ can be computed as
\begin{eqnarray}
\kappa^2((1-\theta)P_S\alpha_{S,R}\|\textbf{h}_{S,R}\|^2+1)T=\theta\eta\alpha_{S,R}P_S\|\textbf{h}_{S,R}\|^2T.\label{eqn8}
\end{eqnarray}

Based on the AF relaying scheme, the signal-to-noise ratio (SNR) at
the destination is given by (\ref{eqn9}) at the top of the next
page,
\begin{figure*}
\begin{eqnarray}
\gamma_D&=&\frac{|\sqrt{\alpha_{R,D}}\textbf{h}_{R,D}^H\sqrt{1-\theta}\textbf{W}\sqrt{P_S\alpha_{S,R}}\textbf{h}_{S,R}|^2}{\|\sqrt{\alpha_{R,D}}\textbf{h}_{R,D}^H\textbf{W}\|^2+1}\nonumber\\
&=&\frac{a\theta(1-\theta)|\textbf{h}_{R,D}^H\hat{\textbf{h}}_{R,D}|^2\|\textbf{h}_{S,R}\|^4}{b\theta|\textbf{h}_{R,D}^H\hat{\textbf{h}}_{R,D}|^2\|\textbf{h}_{S,R}\|^2+\|\hat{\textbf{h}}_{R,D}\|^2(c(1-\theta)\|\textbf{h}_{S,R}\|^2+1)},\label{eqn9}
\end{eqnarray}
\end{figure*}
where $a=\eta P_S^2\alpha_{S,R}^2\alpha_{R,D}$, $b=\eta
P_S\alpha_{S,R}\alpha_{R,D}$ and $c=P_S\alpha_{S,R}$. Similarly, the
received SNR at the eavesdropper is given by (\ref{eqn10}) at the
top of this page,
\begin{figure*}
\begin{eqnarray}
\gamma_E
&=&\frac{e\theta(1-\theta)|\textbf{h}_{R,E}^H\hat{\textbf{h}}_{R,D}|^2\|\textbf{h}_{S,R}\|^4}{f\theta|\textbf{h}_{R,E}^H\hat{\textbf{h}}_{R,D}|^2\|\textbf{h}_{S,R}\|^2+\|\hat{\textbf{h}}_{R,D}\|^2(c(1-\theta)\|\textbf{h}_{S,R}\|^2+1)},\label{eqn10}
\end{eqnarray}
\end{figure*}
where $e=\eta P_S^2\alpha_{S,R}^2\alpha_{R,E}$ and $f=\eta
P_S\alpha_{S,R}\alpha_{R,E}$.

Letting $C_D$ and $C_E$ be the legitimate channel and the
eavesdropper channel capacities, then the secrecy rate is given by
$C_{SEC}=[C_D-C_E]^+$, where $[x]^+=\max(x,0)$ \cite{PLS}. Since
there is no knowledge of the eavesdropper channel at the source and
relay, it is impossible to maintain a steady secrecy rate over all
realizations of fading channels. In this context, we take the
secrecy outage capacity $C_{SOC}$ as the performance metric, which
is defined as the maximum available rate while the outage
probability that the transmission rate surpasses the secrecy rate is
equal to a given value $\varepsilon$, namely
\begin{equation}
P_r(C_{SOC}>C_D-C_E)=\varepsilon.\label{eqn11}
\end{equation}

\section{Optimal Power Splitting}
In this section, we first analyze the secrecy outage capacity of
such an LS-MIMO AF relaying system with energy harvesting, and then
derive an optimal power splitting scheme to determine the
proportional factor $\theta$.

\subsection{Secrecy Outage Capacity}
According to (\ref{eqn11}), the secrecy outage capacity is jointly
determined by the legitimate and eavesdropper channel capacities, so
we first analyze the legitimate channel capacity. Based on the
received SNR at the destination in (\ref{eqn9}), we have the
following theorem:

\emph{Theorem 1}: The legitimate channel capacity in an LS-MIMO AF
relaying system under imperfect CSI can be approximated as
$C_D=W\log_2\left(1+\frac{a\theta(1-\theta)\rho N_R^3}{b\theta\rho
N_R^2+c(1-\theta)N_R+1}\right)$, where $W$ is a half of the spectral
bandwidth.

\begin{proof}
Please refer to Appendix I.
\end{proof}

It is found that the legitimate channel capacity is a constant due
to channel hardening in such an LS-MIMO AF relaying system. Then,
according to the definition of the secrecy outage capacity, we have
a theorem as below:

\emph{Theorem 2}: Given an outage probability bound by
$\varepsilon$, the secrecy outage capacity for an LS-MIMO AF
relaying scheme is
$C_{SOC}=W\log_2\left(1+\frac{a\theta(1-\theta)\rho
N_R^3}{b\theta\rho N_R^2+c(1-\theta)N_R+1}\right)
-W\log_2\left(1+\frac{e\theta(1-\theta)N_R^2\ln\varepsilon}{f\theta
N_R\ln\varepsilon-c(1-\theta)N_R-1}\right)$.

\begin{proof}
Please refer to Appendix II.
\end{proof}

While Theorem 2 is useful to study the secure wireless information
and power transfer system's secrecy outage capacity, the expression
is in general too complex to gain insight. Motivated by this, we
carry out asymptotic analysis on the secrecy outage capacity at high
transmit power regime, and derive the following theorem:

\emph{Theorem 3}: At high transmit power regime, the secrecy outage
capacity in this case is independent of transmit power $P_S$. There
exists a performance upper bound on  the secrecy outage capacity.

\begin{proof}
Please refer to Appendix III.
\end{proof}

\emph{Remarks}: The secrecy outage capacity will be saturated once
the transmit power surpasses a threshold. This is because the AF
relaying system is noise limited due to noise amplification at the
relay in this case. Thus, it makes sense to find the minimum power
that achieves the maximum secrecy outage capacity.

\subsection{Optimal Power Splitting}
According to Theorem 2, for a given transmit power, the secrecy
outage capacity is a function of power splitting ratio $\theta$.
Intuitively, a large $\theta$ leads to a high transmit power at the
relay, but the received signal power decreases. Moreover, from the
view of wireless security, a high transmit power at the relay may
also increases the interception probability. Thus, it is necessary
to optimize the power splitting ratio, so as to maximize the secrecy
outage capacity.

Since $\log_2(x)$ is an increasing function, in order to maximize
the secrecy outage capacity, it is equivalent to maximizing the term
$\frac{1+\frac{a\theta(1-\theta)\rho N_R^3}{b\theta\rho
N_R^2+c(1-\theta)N_R+1}}{1+\frac{e\theta(1-\theta)N_R^2\ln\varepsilon}{f\theta
N_R\ln\varepsilon-c(1-\theta)N_R-1}}$. Thus, the optimal power
splitting can be described as the following optimization problem:
\begin{eqnarray}
\textrm{OP1}:&&
\max\limits_{\theta}\frac{1+\frac{a\theta(1-\theta)\rho
N_R^3}{b\theta\rho
N_R^2+c(1-\theta)N_R+1}}{1+\frac{e\theta(1-\theta)N_R^2\ln\varepsilon}{f\theta N_R\ln\varepsilon-c(1-\theta)N_R-1}}\label{eqn13}\\
\textrm{s.t.}&& 0\leq\theta\leq1.\nonumber
\end{eqnarray}

The objective function is not concave, so it is difficult to provide
a closed-form expression for the optimal $\theta$. However, because
(\ref{eqn13}) is a one-dimensional function of $\theta$, it is
possible to get the optimal $\theta$ by numerical searching.
Specifically, by scaling $\theta$ from 0 to 1 with a fixed step, the
optimal $\theta$ related to the maximum objective function is
obtained.


\section{Numerical Results}
To examine the accuracy and effectiveness of the derived theoretical
results for secure WIPT in an LS-MIMO AF relaying system, we present
several simulation results in the following scenarios: we set
$N_R=100$, $W=10$KHz, $\eta=0.8$, $\theta=0.1$ and $\rho=0.9$
without extra explanation. The relay is in the middle of a line
between the source and the destination. We normalize the path loss
as $\alpha_{S,R}=\alpha_{R,D}=1$, and use $\alpha_{R,E}$ to denote
the relative interception path loss. Specifically, if
$\alpha_{R,E}>1$, the interception distance is shorter than the
legitimate propagation distance, and then the interception becomes
strong. In addition, we use SNR$=10\log_{10}P_S$ to represent the
transmit signal-to-noise ratio (SNR) in dB at the source.

\begin{figure}[h] \centering
\includegraphics [width=0.5\textwidth] {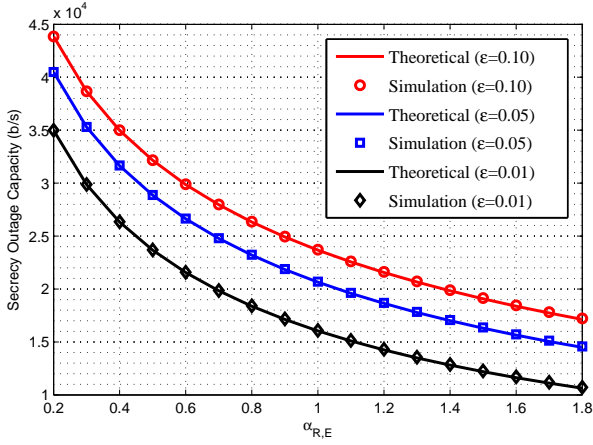}
\caption {Comparison of theoretical and simulation results with
different $\alpha_{R,E}$.} \label{Fig3}
\end{figure}

First, we validate the accuracy of the theoretical results with
SNR=10dB. As shown in Fig.\ref{Fig3}, the theoretical results
coincide with the simulations nicely in the whole $\alpha_{R,E}$
region under different requirements of outage probability. Given a
outage probability bound by $\varepsilon$, as $\alpha_{R,E}$
increases, the secrecy outage capacity decreases gradually. This is
because the interception capacity of the eavesdropper enhances due
to the shorter interception distance. On the other hand, for a given
$\alpha_{R,E}$, the secrecy outage capacity improves with the
increase of $\varepsilon$, since the secrecy outage capacity is an
increasing function of outage probability.

\begin{figure}[h] \centering
\includegraphics [width=0.5\textwidth] {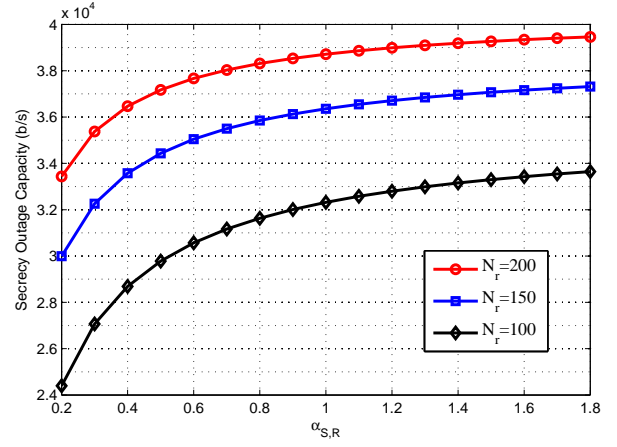}
\caption {Performance comparison with different $\alpha_{S,R}$}
\label{Fig2}
\end{figure}

Second, we investigate the impact of power transfer distance on the
secrecy outage capacity with $\alpha_{R,E}=1$, $\varepsilon=0.01$
and SNR=0dB for secure SWIP. Note that optimal power splitting is
adopted to improve the secrecy outage capacity. As seen in
Fig.\ref{Fig2}, even at a small $\alpha_{S,R}$, namely long transfer
distance, there is a large secrecy outage capacity. As a simple
example, at $\alpha_{S,R}=0.2$, the proposed scheme with $N_r=100$
can achieve $C_{SOC}^{AF}=24$ Kb/s. As $N_r$ increases, the secrecy
outage capacity significantly improves. Thus, the proposed scheme
can solve the challenging problem of long-distance transfer for
secure WIPT.

\begin{figure}[h] \centering
\includegraphics [width=0.5\textwidth] {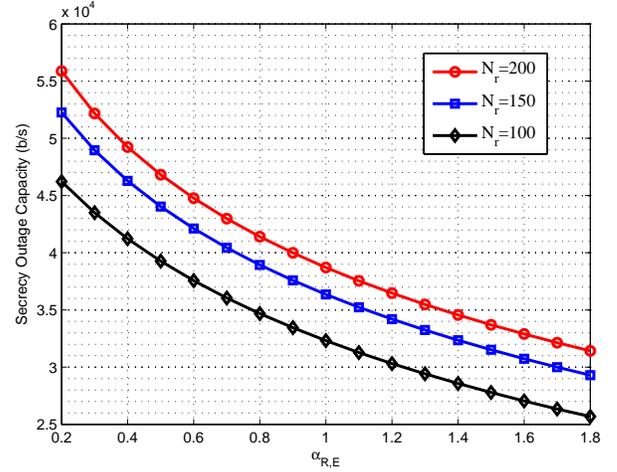}
\caption {Performance comparison with different $\alpha_{R,E}$}
\label{Fig6}
\end{figure}

Then, we examine the effect of interception distance on the secrecy
outage capacity with $\varepsilon=0.01$ and SNR=0dB for secure SWIP.
Similarly, optimal power splitting is adopted to improve the secrecy
outage capacity. As seen in Fig.\ref{Fig6}, even at a large
$\alpha_{R,E}$, namely short interception distance, there is a large
secrecy outage capacity. As a simple example, at $\alpha_{R,E}=1.8$,
the proposed scheme with $N_r=100$ can achieve $C_{SOC}^{AF}=25$
Kb/s. As $N_r$ increases, the secrecy outage capacity significantly
improves. Thus, the proposed scheme can solve the challenging
problem of short-distance interception for secure WIPT.

\begin{figure}[h] \centering
\includegraphics [width=0.5\textwidth] {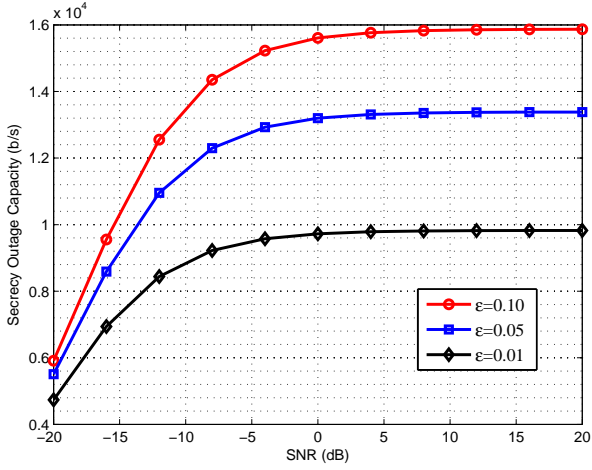}
\caption {Performance comparison with different SNRs.} \label{Fig4}
\end{figure}

Next, we show the impact of SNR on the secrecy outage capacity with
$\alpha_{R,E}=1$. As seen in Fig.\ref{Fig4}, the secrecy outage
capacity is an increasing function of SNR. However, as SNR
increases, the secrecy outage capacity will be saturated. This is
because the AF relaying system is noise limited due to noise
amplification, which confirmed the claim in Theorem 3. Thus, it
makes sense to find the optimal SNR to maximize the secrecy outage
capacity in LS-MIMO relaying systems with the minimum transmit
power.

\begin{figure}[h] \centering
\includegraphics [width=0.5\textwidth] {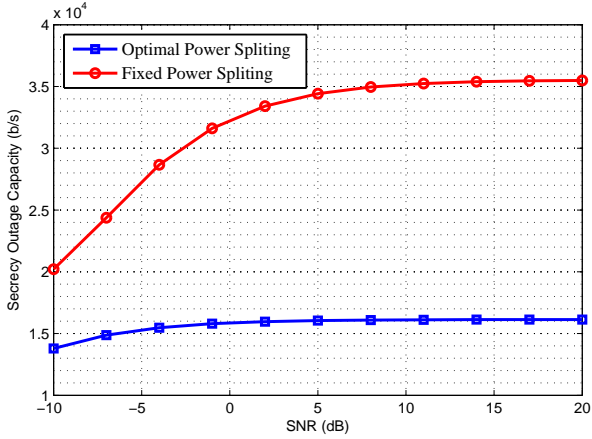}
\caption {Performance comparison with different power splitting
schemes.} \label{Fig5}
\end{figure}

Finally, we check the effectiveness of the proposed optimal power
splitting scheme with respect to a fixed scheme. Specifically, the
fixed scheme sets $\theta$ as 0.1 fixedly regardless of channel
condition and transmit power. As seen in Fig.\ref{Fig5}, the
proposed scheme performs better obviously. Especially, as SNR
increases, the performance gain becomes larger. Note that the
proposed scheme also will suffer from performance saturation.
However, the performance bound can be lifted by adding antennas at
the relay, which is a main advantage of the LS-MIMO relaying scheme
for secure WIPT.

\section{Conclusion}
A major contribution of this paper is the introduction of the
LS-MIMO relaying technique into secure wireless information and
power transfer to significantly enhance wireless security and
improve transmission rate. This paper derives a closed-form
expression of the secrecy outage capacity in terms of transmit SNR,
power splitting ratio, antenna number and interception distance.
Furthermore, through maximizing the secrecy outage capacity, we
present a power splitting scheme, which has a huge performance gain
with respect to the fixed scheme.

\begin{appendices}
\section{Proof of Theorem 1}
Based on the SNR $\gamma_D$ at the destination, the legitimated
channel capacity can be expressed as
\begin{eqnarray}
C\!\!\!\!\!\!\!\!\!&&\!\!\!\!\!\!\!\!\!_{D}\nonumber\\&=&\!\!\!\!W\log_2\bigg(1+a\theta(1-\theta)|\textbf{h}_{R,D}^H\hat{\textbf{h}}_{R,D}|^2\|\textbf{h}_{S,R}\|^4\nonumber\\
&&\bigg/\big(b\theta|\textbf{h}_{R,D}^H\hat{\textbf{h}}_{R,D}|^2\|\textbf{h}_{S,R}\|^2\nonumber\\
&&+\|\hat{\textbf{h}}_{R,D}\|^2(c(1-\theta)\|\textbf{h}_{S,R}\|^2+1)\big)\bigg)\nonumber\\
&=&W\log_2\bigg(1+a\theta(1-\theta)\bigg|(\sqrt{\rho}\hat{\textbf{h}}_{R,D}+\sqrt{1-\rho}\textbf{e})^H\nonumber\\
&&\times\frac{\hat{\textbf{h}}_{R,D}}{\|\hat{\textbf{h}}_{R,D}\|}\bigg|^2
\|\textbf{h}_{S,R}\|^4\bigg/\bigg(b\theta\bigg|(\sqrt{\rho}\hat{\textbf{h}}_{R,D}+\sqrt{1-\rho}\textbf{e})^H\nonumber\\
&&\times\frac{\hat{\textbf{h}}_{R,D}}{\|\hat{\textbf{h}}_{R,D}\|}\bigg|^2\|\textbf{h}_{S,R}\|^2
+(c(1-\theta)\|\textbf{h}_{S,R}\|^2+1)\bigg)\bigg)\label{app1}\\
&=&W\log_2\bigg(1+a\theta(1-\theta)(\rho\|\hat{\textbf{h}}_{R,D}\|^2+2\sqrt{\rho(1-\rho)}\nonumber\\
&&\times\mathcal{R}(\textbf{e}^H\hat{\textbf{h}}_{R,D})
+(1-\rho)\|\textbf{e}\hat{\textbf{h}}_{R,D}^H\|^2/\|\hat{\textbf{h}}_{R,D}\|^2)\|\textbf{h}_{S,R}\|^4\nonumber\\
&&\bigg/\bigg(b\theta(\rho\|\hat{\textbf{h}}_{R,D}\|^2
+2\sqrt{\rho(1-\rho)}\mathcal{R}(\textbf{e}^H\hat{\textbf{h}}_{R,D})\nonumber\\
&&+(1-\rho)\|\textbf{e}\hat{\textbf{h}}_{R,D}^H\|^2/\|\hat{\textbf{h}}_{R,D}\|^2)\|\textbf{h}_{S,R}\|^2\nonumber\\
&&+(c(1-\theta)\|\textbf{h}_{S,R}\|^2+1)\bigg)\bigg)\nonumber\\
&\approx&W\log_2\bigg(1+a\theta(1-\theta)\rho\|\hat{\textbf{h}}_{R,D}\|^2\|\textbf{h}_{S,R}\|^4\nonumber\\
&&\bigg/(b\theta\rho\|\hat{\textbf{h}}_{R,D}\|^2\|\textbf{h}_{S,R}\|^2+c(1-\theta)\|\textbf{h}_{S,R}\|^2+1)\bigg)\label{app2}\\
&\approx&W\log_2\left(1+\frac{a\theta(1-\theta)\rho
N_R^3}{b\theta\rho N_R^2+c(1-\theta)N_R+1}\right),\label{app3}
\end{eqnarray}
where $W$ is a half of the spectral bandwidth, since a complete
transmission requires two time slots. $\mathcal{R}(x)$ denotes the
real part of $x$. $\textbf{h}_{R,D}$ is replaced by
$\sqrt{\rho}\hat{\textbf{h}}_{R,D}+\sqrt{1-\rho}\textbf{e}$ in
(\ref{app1}). (\ref{app2}) follows from the fact that
$\rho\|\hat{\textbf{h}}_{R,D}\|^2$ scales with the order
$\mathcal{O}(\rho N_R)$ as $N_R\rightarrow\infty$ while
$2\sqrt{\rho(1-\rho)}\mathcal{R}(\textbf{e}^H\hat{\textbf{h}}_{R,D})
+(1-\rho)\|\textbf{e}\hat{\textbf{h}}_{R,D}^H\|^2/\|\hat{\textbf{h}}_{R,D}\|^2$
scales as the order $\mathcal{O}(1)$, which can be negligible.
(\ref{app3}) holds true because of
$\lim\limits_{N_R\rightarrow\infty}\frac{\|\hat{\textbf{h}}_{R,D}\|^2}{N_R}=1$
and
$\lim\limits_{N_R\rightarrow\infty}\frac{\|\textbf{h}_{S,R}\|^2}{N_R}=1$,
namely channel hardening \cite{ChannelHardening}. Therefore, we get
the Theorem 1.

\section{Proof of Theorem 2}
According to (\ref{eqn11}), for a given $\varepsilon$, we have
\begin{eqnarray}
\varepsilon&=&P_r\left(C_{SOC}>C_{D}-W\log_2(1+\gamma_{E})\right)\nonumber\\
&=&P_r\left(\gamma_E>2^{\left(C_{D}-C_{SOC}\right)/W}-1\right)\nonumber\\
&=&1-F\left(2^{\left(C_{D}-C_{SOC}\right)/W}-1\right),\label{app4}
\end{eqnarray}
where $F(x)$ is the cumulative distribution function (cdf) of
$\gamma_E$. In order to derive the secrecy outage capacity, the key
is to get the cdf of $\gamma_E$. Examining (\ref{eqn10}), due to
channel hardening, we have
\begin{eqnarray}
\gamma_E=\frac{e\theta(1-\theta)N_R^2\left|\textbf{h}_{R,E}^H\frac{\hat{\textbf{h}}_{R,D}}{\|\hat{\textbf{h}}_{R,D}\|}\right|^2}
{f\theta
N_R\left|\textbf{h}_{R,E}^H\frac{\hat{\textbf{h}}_{R,D}}{\|\hat{\textbf{h}}_{R,D}\|}\right|^2+c(1-\theta)N_R+1}.\label{app5}
\end{eqnarray}
Since $\hat{\textbf{h}}_{R,D}/\|\hat{\textbf{h}}_{R,D}\|$ is an
isotropic unit vector and independent of $\textbf{h}_{R,E}$,
$\left|\textbf{h}_{R,E}^H\hat{\textbf{h}}_{R,D}/\|\hat{\textbf{h}}_{R,D}\|\right|^2$
is $\chi^2$ distributed with 2 degrees of freedom. Let
$y\sim\chi_2^2$, we can derive the cdf of $\gamma_E$ as
\begin{eqnarray}
F(x)&=&P_r\left(\frac{e\theta(1-\theta)N_R^2y}{f\theta
N_Ry+c(1-\theta)N_R+1}\leq x\right).\nonumber\\\label{app6}
\end{eqnarray}
If $x<e(1-\theta)N_R/f$, then we have
\begin{eqnarray}
F(x)&=&P_r\left(y\leq\frac{(c(1-\theta)N_R+1)x}{e\theta(1-\theta)N_R^2-f\theta N_Rx}\right)\nonumber\\
&=&1-\exp\left(-\frac{(c(1-\theta)N_R+1)x}{e\theta(1-\theta)N_R^2-f\theta
N_Rx}\right).\nonumber\\\label{app7}
\end{eqnarray}
Since $x\geq e(1-\theta)N_R/f$ is impossible when
$x=2^{\left(C_{D}-C_{SOC}\right)/W}-1$, we have
\begin{equation}
\varepsilon=\exp\left(-\frac{(c(1-\theta)N_R+1)x}{e\theta(1-\theta)N_R^2-f\theta
N_Rx}\right).\label{app8}
\end{equation}
Equivalently, we have
\begin{eqnarray}
C_{SOC}&=&W\log_2\left(1+\frac{a\theta(1-\theta)\rho
N_R^3}{b\theta\rho
N_R^2+c(1-\theta)N_R+1}\right)\nonumber\\
&&-W\log_2\left(1+\frac{e\theta(1-\theta)N_R^2\ln\varepsilon}{f\theta
N_R\ln\varepsilon-c(1-\theta)N_R-1}\right).\nonumber\\\label{app9}
\end{eqnarray}
Hence, we get the Theorem 2.

\section{Proof of Theorem 3}
For the secrecy outage capacity in (\ref{app9}), if the transmit
power $P_S$ is sufficiently large, it can be approximated as
\begin{eqnarray}
C\!\!\!\!\!\!\!\!\!&&\!\!\!\!\!\!\!\!\!_{SOC}\nonumber\\&\approx&W\log_2\left(\frac{a\theta(1-\theta)\rho
N_R^3}{b\theta\rho
N_R^2+c(1-\theta)N_R}\right)\nonumber\\
&&-W\log_2\left(\frac{e\theta(1-\theta)N_R^2\ln\varepsilon}{f\theta
N_R\ln\varepsilon-c(1-\theta)N_R}\right)\label{app10}\\
&=&W\log_2\left(\frac{\eta
P_S\alpha_{S,R}\alpha_{R,D}\theta(1-\theta)\rho
N_R^2}{\eta\alpha_{R,D}\theta\rho N_R+1-\theta}\right)\nonumber\\
&&-W\log_2\left(\frac{\eta P_S\theta\alpha_{R,E}
N_R\alpha_{S,R}(1-\theta)\ln\varepsilon}{\eta\theta\alpha_{R,E}\ln\varepsilon-(1-\theta)}\right)\nonumber\\
&=&W\log_2\left(\frac{\alpha_{R,D}N_R(\eta\theta\alpha_{R,E}\ln\varepsilon-(1-\theta))}
{(\eta\alpha_{R,D}\rho\theta
N_R+(1-\theta))\alpha_{R,E}\ln\varepsilon}\right),\label{app12}
\end{eqnarray}
where (\ref{app10}) holds true since the constant ``1" is negligible
when $P_S$ is large enough. It is found that the secrecy outage
capacity in this case is independent of transmit power $P_S$. Thus,
we prove the Theorem 3.

\end{appendices}

\end{document}